\documentclass[3p,times,twocolumn,leqno]{elsarticle}

\usepackage{ecrc}
\usepackage{ecrc}
\usepackage{graphicx}
\usepackage{epstopdf}

\volume{00}
\firstpage{1}

\journalname{Nuclear Physics B Proceedings Supplement}
\runauth{}
\jid{nuphbp}
\jnltitlelogo{Nuclear Physics B Proceedings Supplement}
\usepackage{amssymb}
\usepackage[figuresright]{rotating}

\begin{document}
\begin{frontmatter}

\dochead{}

\title{Neutrino oscillations along the Earth to probe flavor parameters: a \emph{NeuWorld}}

\author{Daniele Fargion  and Daniele D'Armiento}
\address{Physics Dept. University Rome University, Sapienza; Ple.A.Moro 5,00185,Rome,Italy; INFN Rome 1}

\begin{abstract}
On 2010 MINOS experiment was showing an hint of possible different mass splitting and mixing angles for neutrinos and anti-neutrinos, suggesting a charge-parity-time (CPT) violation in the lepton sector;
 later on last year 2012 a second result from MINOS showed a reduced discrepancy between the two set of parameters, nearly compatible with no CPT
violation.  We proposed an experiment for more precise estimation of neutrino and $\nu - \bar{\nu}$ oscillation parameters being useful not only for
a complete discrimination for CPT scenarios, but also for mass hierarchy and $\theta_{13}$ determination, and mostly for the first oscillated detection of tau , $\nu_{\tau} $ and $ \bar{\nu_{\tau}}$, neutrinos.
Indeed, the last a few years of OPERA activity on the appearance of a $\tau$ lepton still
has not been probed convincingly. Moreover atmospheric anisotropy
in muon neutrino spectra in IceCube DeepCore, at $\cong 10$ GeV, can hardly reveal asymmetry in the
$\nu_{\mu} - \bar{\nu_{\mu}}$ oscillation parameters nor the tau and anti tau appearance.
We show an experiment, (the longest baseline neutrino oscillation test available by
crossing most of Earth's diameter, a \emph{NeuWorld}), that may improve the oscillation measurement and disentangle at best any hypothetical CPT
violation  while testing $\tau$ and $\bar{\tau}$ appearance at the same time of the $\nu_{\mu} - \bar{\nu}_{\mu}$ disappearance correlated.
The experiment use a neutrino beam crossing the
Earth, within an OPERA-like experiment from CERN (or Fermilab), in the direction of the IceCube DeepCore
detector at the South Pole.  Such a tuned beaming and its detection  may lead to a strong signature of neutrino muon-tau mixing (nearly one $\bar{\tau}$  or two $\tau$ a day, with $10 \sigma$ a year), even for an one per-cent OPERA-like experiment.
\end{abstract}

\end{frontmatter}

\section{Introduction}
The MINOS 2010 observations  \cite{1} seemed to imply (or now
at least to marginally hint \cite{111})
  a different anti-neutrino mass splitting with respect to well known neutrino one, leading to a possible CPT violation.
   Even a marginal neutrino anti-neutrino mass difference may open new roads in our particle
  physics understanding \cite{4}.
  This CPT violation might indicate a very peculiar role of neutral leptons in matter/anti-matter
  genesis, and it may address unsolved lepton-baryon-genesis open puzzle, related to cosmological mysteries. Consequently
  such a tiny CPT violation, if confirmed, might become one or the main (amazing) discovery of the century: therefore we inquire how at  best
CPT violation may be observed. In order to do that we proposed
two ways: (1) exploiting current and coming soon atmospheric data
from IceCube DeepCore detector, and (2) considering a completely
new neutrino oscillation baseline. Such experiment should beam neutrinos across the whole Earth: \emph{NeuWorld}\cite{-1}, and it is almost ready since
both accelerator and detector already exist suitable for that. Only a new tunnel, Opera-like,  is necessary.

\section{A road map to disentangle mixing flavors}
\begin{figure}[hbt]
\includegraphics[scale=.12]{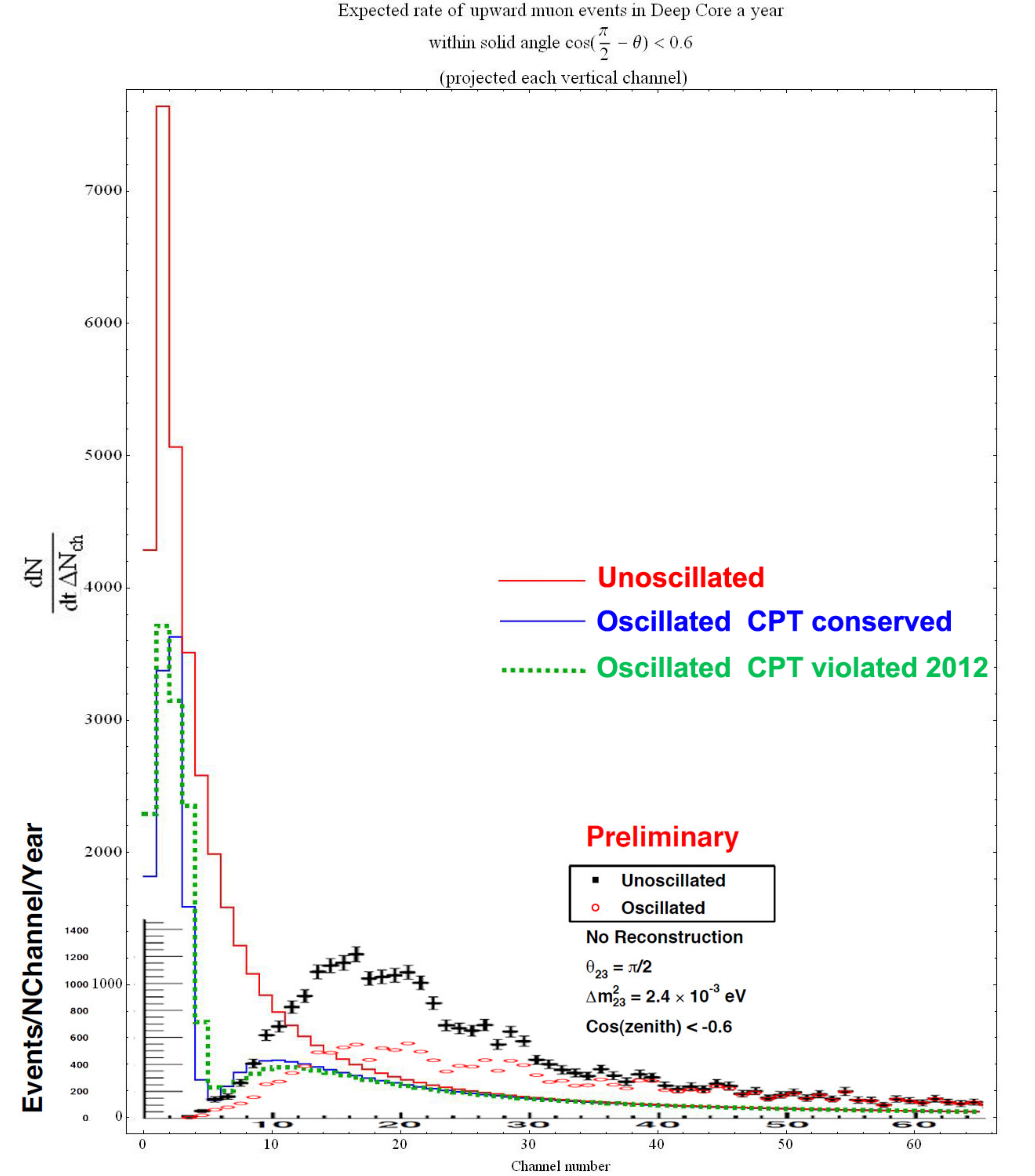}
\caption{Our expected atmospheric neutrino spectra in DeepCore, as a function of the DOM channel number, for the latest (2012) neutrino oscillation
parameters. Here we apply the Earth-matter influence along the overall oscillation vs. the DeepCore preliminary expectations ($2009\ -\ 2010$),\cite{0,7}. The
green dotted curve (oscillation in case of CPT violation) is very close to the CPT-conserved rate denoted by
the thin blue line. They are to be compared to DeepCore small red ovals estimation, by \cite{0}, which has been used and
shown here as a reference benchmark. While the CPT deviation in 2010 was, in principle, well detectable because of a $30\%$ suppression (with respect to the CPT-conserved case, \cite{-1})
 around channels 8 - 12, now a small CPT deviation, as  in 2012 MINOS data, is almost unobservable. Note that our predictions rate shown by blue and green dots at channels above 20 (corresponding after geometry projection to nearly 70-80 GeV  neutrino energy)  are nearly half of the
 corresponding ones in Deep Core foreseen event rate (shown as small red circles). This discrepancy might be observable as soon data will be released.}\label{Fig_1}
\end{figure}

We considered first the ongoing experiment based on atmospheric
neutrino signal in DeepCore that may be a benchmark for CPT
scenario disentangling; we underline that the smeared muon track energy measure and
the angular resolution can confuse any tiny CPT
anisotropy in Deep Core spectra (Fig \ref{Fig_1}); such smeared anisotropies might be not
 clearly detected  by present atmospheric neutrino records in Deep Core.
Only a negligible difference would rise in the low channel muon tracks
by allowable CPT violation, as shown in Fig. \ref{Fig_1}. On the contrary
the common  muon disappearance (as the SK observation)
will be anyway observable.

Therefore we  focus here on a future possible ad-hoc Very Long
Baseline experiment able to sharply confirm even tiny CPT violation
in a very short time and in an accurate way. We studied the
appearance of muon neutrino by $\nu_{\mu} \rightarrow
\nu_{\mu}$ and ${\bar\nu}_{\mu} \rightarrow {\bar\nu}_{\mu}$
oscillations, in the energy-distance range where CPT conserved
oscillation is almost vanishing, while CPT violated oscillation
is (partially) allowed, based on the experimental parameters
determined by MINOS. By doing this we will be able also to test
with great accuracy the $\nu_{\mu} \rightarrow \nu_{\tau}$, and ${\bar\nu}_{\mu} \rightarrow {\bar\nu}_{\tau}$
mixing, leading also to a very precise estimate of their mixing
parameters that may shed also light to a possibly hidden symmetry
unwritten into a tuned value: $\sin(2 \theta_{23})\simeq 1$.

We performed an estimation for other experiment  such as OPERA, \cite{-2} at Gran Sasso, for calibration.
A common shortcoming of most baseline experiments (thousand or hundreds km $\nu$ flight)  is the very un-efficient flavour conversion
probability and the usual low neutrino energy. The higher the energy, the
larger the distance needed to complete a mixing oscillation,
but also the better the  beaming, (because of the higher
Lorentz factor of pion decay), as well as the larger the
neutrino cross-section. Incidentally the approximate beaming
solid angle shrinks by a factor proportional to $E_{\pi}^{2}$,
and the neutrino-matter cross-section grows as $E_{\nu}$
providing a global signal enhancement amplified by a factor
proportional to $\sim E_{\nu}^{3}$. Therefore a long-baseline
experiment for instance as in our $NeuWorld$\cite{-1} experiment at 22 GeV,  may play
a better role ($8000$ times better than 1 GeV experiment) to define oscillation parameters. Moreover the dilution
factor due to the much greater distance from CERN of DeepCore
than OPERA  \cite{-2}, a factor $\simeq 240$, is widely compensated by the
detector mass ratio (DeepCore  versus OPERA), at least by a
factor $\simeq 4800$, implying a benefit of a factor  $\simeq
20$. In addition the larger distance in the  $NeuWorld$\cite{-1}  baseline
offers a complete $\nu_{\mu}\leftrightarrow \nu_{\tau}$
conversion with respect to $1.5 \%$ of OPERA, providing a
further gain of an additional factor $\simeq 60$. All together
the advantages of a long baseline experiment with DeepCore
(respect to OPERA) in tau appearance,  is a factor of about or
above $2400$; moreover all the born $\tau$  (within the limited
4.8 Mton DeepCore) will be observable (contrary to OPERA), leading to
an efficiency ratio  $15:1$ for
$NeuWorld$, leading to an exceptional ratio $36000$ between $NeuWorld$ and OPERA  \cite{-2} in testing  tau appearance.
One tau a day in our scenario even for just at $1\%$ OPERA size $NeuWorld$  experiment versus one rare
tau a year in present OPERA experiment (see more precise
details in next Tables). We remind that we are considering half
detection volume respect  the one claimed, just for prudential
reasons \cite{15,12}.\\

Our main proposal therefore stands for building a baseline from
CERN (or FNAL) to IceCube DeepCore detector, at $E_{\nu} \sim
20 $ GeV, with more than 4 MegaTon detection mass.
SuperKamiokande detector was also considered, but found to be
not enough in terms of signal intensity, in comparison to
DeepCore.

Therefore we estimated in detail, for each source (CERN or
FNAL), for each Proton On Target source
 flux (and its secondary neutrino flux) and for each detector distance (SK or IceCube), the results
 (muons or tau events) by a chain of  neutrino-signal values
source-propagation-flavor mixing and oscillation in Earth, the
detection rate in volume inside or outside the detector.  Each
value or formula is deeply correlated to the previous one,
leading to a realistic estimate of muon or anti-muon signals
(as well as tau-anti tau), designed  at best to disentangle the
hypothetical CPT asymmetry as well as to fix with high accuracy
muon-tau flavor mixing parameters.

\section{\emph{NeuWorld}:Neutrino events in longest baselines}
\begin{figure}[hbt]
\includegraphics[scale=.27]{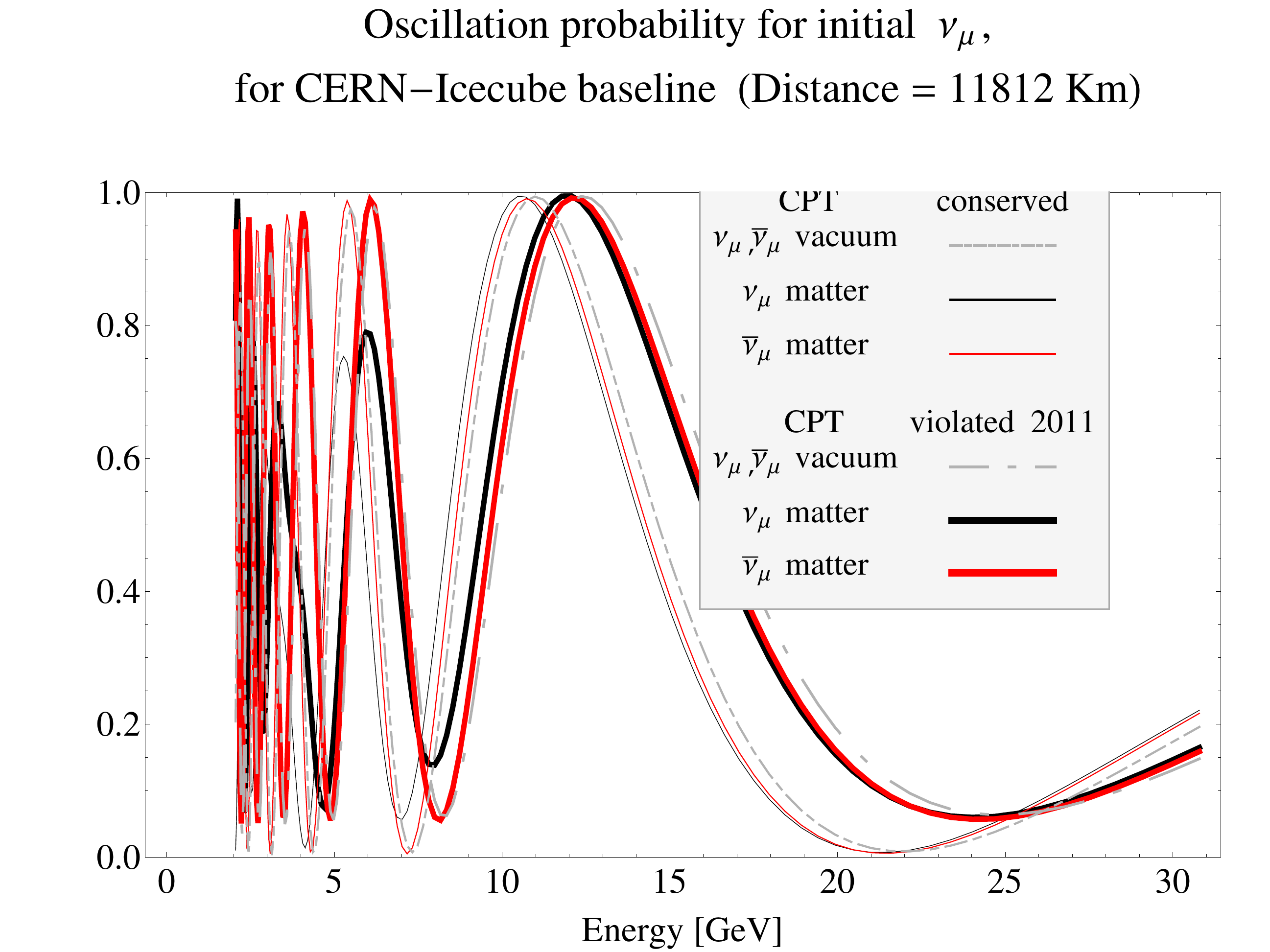}
\caption{Muon neutrino and anti-neutrino survival probability ($ P_ {\mu \mu}$ and  $P_ {\bar{\mu} \bar{\mu}}$).
There is a difference due to a slight asymmetry in the MSW term for matter effect;
moreover such a discrepancy between matter and anti-matter is negligible in
our energy window of interest. The FNAL-SK and FNAL-IceCube oscillations
are very comparable to those from CERN because of very similar distances. Note that while crossing
the Earth here and later we did not consider (as most other authors did) only the
average density of Earth (respectively, $\rho = 4.5 g cm^{-3} $ for "near" SK; $\rho = 7.2
g cm^{-3} $ for "far" IceCube DeepCore), but the exact variable matter profile.
 Nevertheless, the vacuum and matter cases, as shown in figure,  do not differ relevantly
 for energies higher than 10 GeV.}\label{Fig_3}
\end{figure}

In figure \ref{Fig_3} we show muon neutrino oscillation
probability versus neutrino energy, for the two set of
parameters, conserving and violating CPT, that is assuming the
same parameters for both $\nu - \bar{\nu}$ (under CPT conserved case) and
different parameters between $\nu$ and  $\bar{\nu}$ (under CPT
violated case). It is clearly visible that in the eventual CPT
violated scenario, a different oscillation for $\nu_{\mu}$ and
$\bar{\nu}_{\mu}$ would result, leading to a discrepancy in the
event number, while this would not be for usual symmetry
between matter and anti-matter. Therefore we suggest this new
$\nu_{\mu} \rightarrow \nu_{\mu}$ and ${\bar\nu}_{\mu} \rightarrow {\bar\nu}_{\mu}$
 disappearance experiment, and propose various longest
baselines, for which both source and detector are already
available. All considered baselines are listed in table 1.

The method of estimating the beaming and detection
of neutrinos to detectors (DeepCore or SuperKamiokande)
is based on a chain of correlated evaluations
that we used and calibrated with known experiments:
OPERA and MINOS.
By evaluating the chord distance and the appropriate energy enhancing neutrino
oscillation, we find $\nu_{\mu}$ event rate as would be no oscillation, in comparison to known rates in OPERA for instance,
using the same p.o.t. number (table 1). The energy is chosen as to maximize $\nu_{\mu}$
 disappearance, so that any eventual different $\bar{\nu}_{\mu}$  oscillation probability (in CPT violated case) would
 be clearly visible over a nearly null background. We do remind that because accurate timing
 in sending neutrino bunches along the Earth one
 may easily cancel any rare upward atmospheric neutrino noise,
also because  of the selected narrow angle arrival direction.

We always take in consideration
the flavor neutrino mixing within the Earth, keeping care of the exact step by step
variability of the inner matter density. At
energies $ \geq 20$ GeV, nevertheless, the flavor mixing along
the Earth diameter is not very much different from the vacuum
case, but we did take into account of it.

In table 2 we consider the primary bunch bending and tunnel parameters, for a 1 Tesla magnetic field curving a pion
beam at nearly 50 GeV energy. In the most economic version
the whole tunnel arc with the rectilinear sector length  will be less than 200 m  underground.
This tunnel cost is the only expenses  needed to invest to achieve a the \emph{NeuWorld} experiment.
Detectors (Deep Core) and accelerators (CERN or FNL) already exist.

In figure 3 it is shown how oscillation probability is
averaged by  a non-monochromatic
$(\Delta E / E =  20\%) $ neutrino beam. The presence
of such energy smearing increases the noise and reduces the
signal significance. This severe noise is usually ignored by most of the other author papers.
However, as discussed in the following
tables, the event rate in the worst $1\%$ OPERA-like experiment
may lead to a remarkable $6 \sigma $ signal detection of any tiny (MINOS 2012) hypothetical
CPT violation within a year of detection.

In table 3  we take in account the averaged oscillation probability and obtain
the resulting event for $\nu_{\mu}$ and  $\bar{\nu}_{\mu}$;  the latter  $\bar{\nu}_{\mu}$ is described both for conserved or violated CPT scenarios.
These events are compared to a OPERA-like experiment at  $100\%$ (decay length and neutrino flux) , while
in table 4 it is shown  a severe economic reduction  $1\%$ because a $5\%$  shorter pion decay length and additional $20\%$ due to considering
a lower $\nu_{\mu}$ intensity flux, since our proposed beam select pion flux by a spectrometer energy filter while bending the bunch.

The (preliminary) correlated parameter map derived
by a year of recording (in a minimal $1\%$ OPERA beaming
experiment) is shown in Figure 4.\\

We also underline the $\nu_{\tau}$ appearance, so that our experiment \emph{NeuWorld}
has another important return: this will be able to detect nearly 2 $\tau$ a day, or 1 $\bar{\tau}$
as seen in table 5 and 6. The noise of Neutral Current events was considered as background,
leading nevertheless to a $14$ $ \sigma$ for tau appearance and $10$ $ \sigma$ for anti-tau appearance.

%\begin{figure}[hbt]
%\includegraphics[scale=.3]{polosud-24_6.pdf}
%\caption{ Neutrino and anti-neutrino Muon  oscillation probability in ($\nu_{\mu}\rightarrow\nu_{\tau} $), ($\overline{\nu}_{\mu}\rightarrow\overline{\nu}_{\tau} $) oscillation probability for three cases: a first conserving CPT symmetry,  $\sin(2\theta_{23})=\sin(2\overline{\theta}_{23})=1$ and  $\Delta m_{23}^{2}=\Delta \overline{m}_{23}^{2}=(2.35^{+0.11}_{-0.08})\cdot 10^{-3}eV$ (CPT conserved case), a second one violating CPT symmetry \cite{1}, where its anti muon neutrino CPT parameters are: $\Delta \overline{m}_{23}^{2} =(3.36^{+0.45}_{-0.40}($stat.$)\,\pm\,0.06($syst.$))\cdot 10^{-3}$ eV$^{2}$ and $\sin^{2}(2\overline{\theta}_{23})\,=\,0.86\,\pm\,0.11($stat.$)\,\pm\,0.01($syst.$)$, and a third (last) CPT symmetry \cite{111} where: $\Delta \overline{m}_{23}^{2} = 2.62 \cdot 10^{-3}eV$ and $\sin^{2}(2\overline{\theta}_{23})\,=\,0.945$.
%These values are applied along all the article. These cases are shown only here in the simplest vacuum approximation (whose results at high energy are nearly coincident with the matter ones to be applied later).  At bottom figure the lowest $24.6$ GeV case shows  a clear variability and anisotropy due to the muon disappearance to be  observed in DeepCore.  An average anisotropy at tens GeV maybe detected as in next figures.  The vertical dotted lines refer to the baseline distances for OPERA, SuperK and ICECUBE, or across all the Earth as in Table 1. It is to note the very small conversion probability at the OPERA distance.}\label{Fig_2}
%\end{figure}
%\begin{landscape}

\begin{table}[htbp]
%\begin{center}
%\quad\\
\begin{tiny}
\begin{tabular}{lccccc}
\hline
\hline
&&&&&\\
\multicolumn{6}{c}{Muon Neutrino beam events by}\\
\multicolumn{6}{c}{3.5 $\cdot 10^{19}$ proton on target(p.o.t) a year from CERN or FNAL }\\
&&&&&\\
\hline
Baseline & distance &  $E_{\nu}$ & $\left(\frac{L'}{L}\right)^2$ & Mass detector  & $N_{ev \mu_{CC}} $ no osc. \\
&(km) & (GeV)  & & $kton$ &  $ year^{-1}$\\
\hline
CERN--OPERA&$        L =732$ &    $17$  &   1 &      $ 1.2 $   & 2847     \\
CERN--SK&$            L' =8737$& $15.8$  & $142.5$ & $ 22.5$  &   324   \\
Fermilab--SK&$       L' =9140$& $16.5$  &  $155.9$ & $ 22.5$  &   341   \\
CERN-IceCube&$       L' =11812$& $21.8$  & $260.4$ & $ 4800$  &   93343  \\
Fermilab--IceCube&  $L' =11623$& $21.4$ &  $252.1$ & $ 4800$  &   93951   \\
\hline
\end{tabular}
\end{tiny}
%\end{center}
\caption{ Source detector distances, tuned energies, flux
dilution, event rate, in comparison with the OPERA
baseline.}\label{f}
\end{table}
%\end{landscape}

\begin{table}[htbp]
\begin{scriptsize}
\begin{tabular}{lccccc}
\hline
\hline
Baseline & $ E_{\nu}$ &    Angle      & Arc    &   Tunnel       &   Tunnel   \\
         &            &               &     Depth & length  &       depth         \\
         &   (GeV)    &   (deg)      &  (m)      &  $L_{5\%}$ (m) & $H_{5\%}$ (m) \\
\hline
&&&&&\\
CERN--SK&            $15.8$&  $ 43.19°$   & $ 34.2$  & $50$ &34     \\
Fermilab--SK&        $16.5$&   $45.77°$    &$ 40  $  & $55$ &40     \\
CERN-IceCube&        $21.8$&   $67.82°$    &$ 106.3$  & $92$ &85     \\
Fermilab--IceCube&   $21.4$&   $65.67°$    &$ 99.5$  & $90$ &82     \\
\hline
\end{tabular}\\
\caption{Beaming, bending and tunnel parameters: final neutrino energy, bending angle (for 1T magnetic field),  beam bending arc depth, decay tunnel length and depth in an economic scenario, considering shorter  pion decay tunnel length, that is $5\%$ decay length respect the OPERA one (1 km), for the same neutrino flux.}\label{tunnell}
\end{scriptsize}
\end{table}

\begin{figure}[hbt]
\includegraphics[scale=.3]{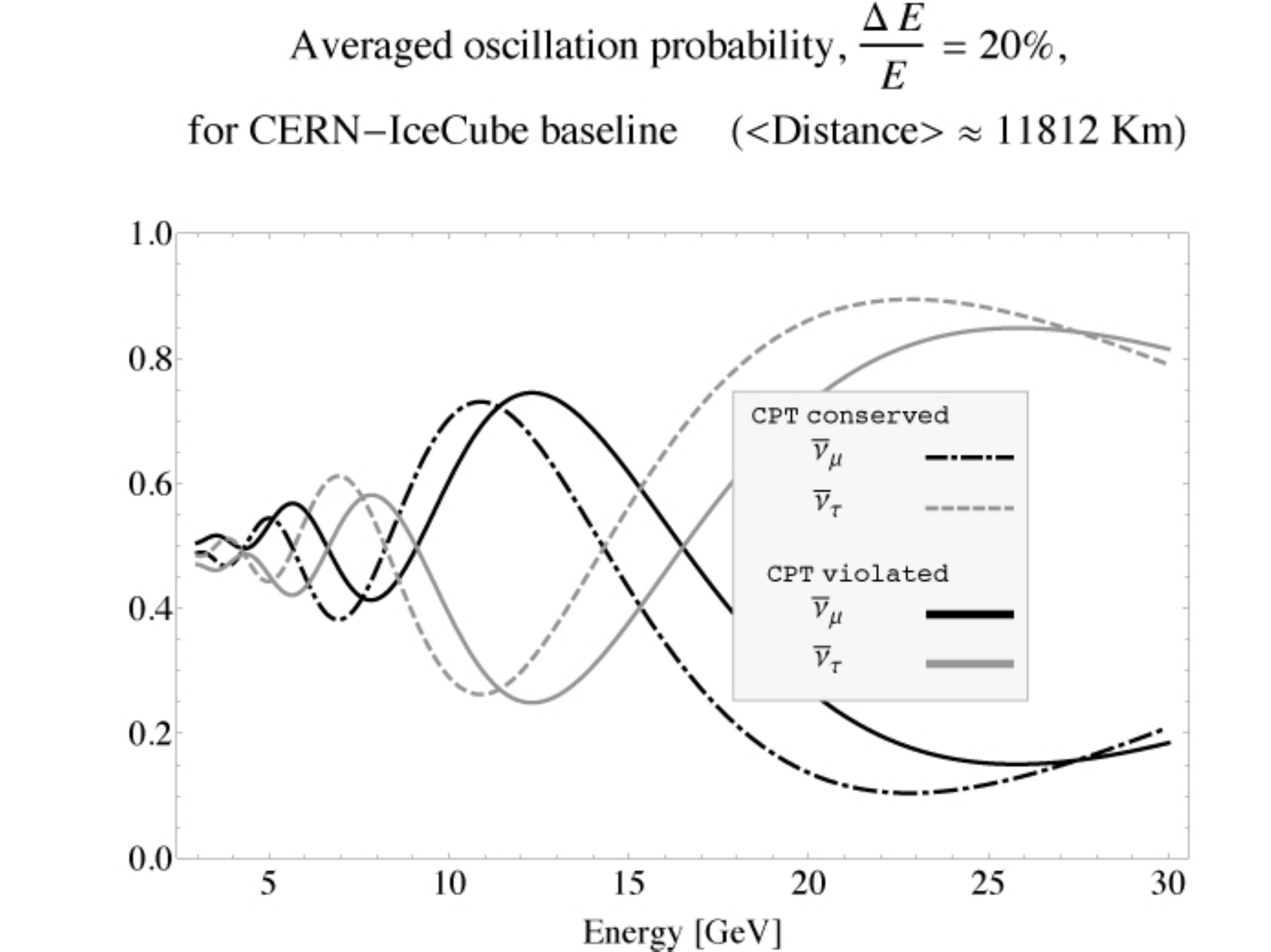}
\caption{Averaged oscillation and survival probability, showing the average
conversion smeared by neutrino energy spectra at $\Delta E/E = \pm 20\% $. The survival
probability is less sharp than in the monochromatic scenario, however,
as shown in Table 4, the event rate even in the framework of a minimal $1\%$
OPERA-like \emph{NeuWorld} experiment allows us to detect any eventual CPT violation with $6.1 \sigma $  signature in one year.}\label{Fig_4}
\end{figure}

\begin{table}[htbp]
\begin{center}
\begin{tiny}
\begin{tabular}{lccccc}
\hline
\hline
Baseline  & $N_{\nu_{\mu}}$ / $N_{\bar{\nu}_{\mu}}$  & $\langle P_{\mu \mu} \rangle$ / $\langle P_{\bar{\mu} \bar{\mu}}\rangle$ & $N_{\nu_{\mu}}$   & $N_{\bar{\nu}_{\mu}}$  & $N_{\bar{\nu}_{\mu}}$  \\
    &  no osc.   & $\mathbf{\frac{\Delta E}{E} = 20 \%}$  &   after osc.  & after osc. & after osc.  \\
    & inside and outside  & $\exists$ CPT  / $\nexists$ CPT    & $\exists$ CPT & $\exists$ CPT & $\nexists$ CPT   \\
\hline
CERN--OPERA     & $21125$   / $10562$  &0.985 /  0.972  & 20494 & 10247  & 10125\\
&&&&\\
CERN--SK        & $1945$  / $972$    & 0.096 /   0.18 & 187 & 93 & 175\\
Fermilab--SK     & $2050$  / $1025$   & 0.096 /  0.179 & 197 &98 & 183\\
CERN--IceCube     & $116679$ / $58340$  &  0.13 /  0.26 & 12415  & 6207& 15168 \\
Fermilab--IceCube & $117439$ / $58720$  &  0.129 /  0.263 & 12496 & 6248 & 15443 \\
\hline
\end{tabular}
\end{tiny}
\caption{$\nu_{\mu}$ and $\bar{\nu}_{\mu}$ appearance: event rate in conserved and in violated CPT
parameters (100\% tunnel length comparable with OPERA).
The events are relative to one year data taking, as they would be without neutrino oscillation.
From the first column: the total number of $\nu_{\mu}$ and $\bar{\nu}_{\mu}$
events born inside detector and those whose track starts outside detector; survival probability for $\nu_{\mu}$ and $\bar{\nu}_{\mu}$ under $20\%$ energy uncertainty;
final number of $\nu_{\mu}$ and $\bar{\nu}_{\mu}$ events for the two CPT scenario.}\label{f3}
\end{center}
\end{table}

\vspace{-0cm}

\begin{table}[htbp]
\begin{scriptsize}
\begin{tabular}{lccccc}
\hline
\hline
 \textbf{1\% }&  $ N_{\nu_{\mu}}$& $N_{\bar{\nu}_{\mu}}$ & $N_{\bar{\nu}_{\mu}}$  & Statistical \\
        &  after osc.  &  after osc.   &   after osc.   & Significance \\
Baseline&  $\exists$ CPT &   $\exists$ CPT   &  $\nexists$ CPT  & $\sigma$ \\
\hline
CERN--SK       &     $1.9$     & $\mathbf{0.9}$   & $\mathbf{1.8\pm1}$  & 0.6  \\
Fermilab--SK    &   $2$      & $\mathbf{1}$       & $\mathbf{1.8\pm1}$  & 0.6  \\
CERN--IceCube  &     $152$     & $\mathbf{76}$    & $\mathbf{152\pm12}$  & $\mathbf{6.1} $  \\
Fermilab--IceCube&   $151$     & $\mathbf{76}$    & $\mathbf{154\pm12}$  &$ \mathbf{6.3}$   \\
\hline
\end{tabular}
\caption{$\nu_{\mu}$ and $\bar{\nu}_{\mu}$ event rates considering reduced size experiment, and $\frac{\Delta E}{E} = 20 \%$ energy spread.
It is shown the very economic setup where event rate is reduced to $1\%$, having considered $5\%$ decay length tunnel, and $20\%$  $\pi$ flux intensity in \emph{NeuWorld} experiment.} \label{f7}
\end{scriptsize}
\end{table}

%TAU
\begin{table}[htbp]
\begin{tiny}
\begin{tabular}{lccccc}
\hline
\hline
 & $\langle P_{\mu \tau} \rangle$  &  $N_{\nu_{\tau}}^{CC}$  & $N_{\bar{\nu}_{\tau}}^{CC}$   & $N_{\nu_{\tau}}^{CC}+N_{\nu_{i}}^{NC} $  &  $N_{\bar{\nu}_{\tau}}^{CC}+N_{\bar{\nu}_{i}}^{NC}$ \\
Baseline &   $\mathbf{\frac{\Delta E}{E} = 10 \%}$  &  {with osc.}  &  {with osc.} \\
\hline
CERN--OPERA       &  $0.015$ &  16   & 0.5  & 16 & 0.5\\
&&&\\
CERN--SK          &  $0.962$  &  119 & 59      & $220$   & $110$    \\
Fermilab--SK      &  $0.963$  &  125  & 63      & $232$   & $116$   \\
CERN--IceCube     &  $0.945$  &  36166 & 18083   & $65289$ & $32645$ \\
Fermilab--IceCube &  $0.944$  &  36363 & 18181  & $65676$ & $32838$ \\
\hline
\end{tabular}
\caption{Tau-AntiTau neutrinos in matter by CPT
conserved-violated case.  Estimated total events of charged
current muon neutrino interaction in detector, conversion
probabilities in matter, cross section ratio between
tau-neutrino and muon neutrino, event rates for tau and
anti-tau neutrino (last with both conserved and violated
CPT-symmetry), and noise events by Neutral Current neutrino interactions. Note that the absence of  $\tau$ or $\bar{\tau}$  appearance would lead to less than half of the event showering rate in last columns.}\label{t1a}
\end{tiny}
\end{table}

\begin{table}
\begin{footnotesize}
\caption{Tau-AntiTau neutrinos event rates for reduced \emph{NeuWorld} experiment, as before, to overall $1\%$ of OPERA experiment.
Statistical significance is referred both for $\nu_{\tau}$, $\bar{\nu}_{\tau}$ only detection, and for CPT cases detection.
Note that the absence of  $\tau$ or $\bar{\tau}$  appearance would lead to less than a half of the showering event rate.}\label{t1a}
\begin{tabular}{lccccc}
\hline
\hline
\textbf{$1\%$ }  &$N_{\nu_{\tau}}^{CC}+N_{\nu_{i}}^{NC} $ &  $\sigma $  & $N_{\bar{\nu}_{\tau}}^{CC}+N_{\bar{\nu}_{i}}^{NC}$ & $\sigma $  \\
Baseline & $\exists$  CPT & for $\nu_{\tau}$ & $\exists$  CPT   & for $\bar{\nu}_{\tau}$  \\
        &  $year^{-1}$    &           &  $year^{-1}$  &    \\
%\cline{1-5} \cline{7-8}
\hline
CERN--SK          & 2   & 0.8 &  1.1                 & 0.6      \\
Fermilab--SK      & 2   & 0.8 &  1.2                 & 0.6      \\
CERN--IceCube     & 653 & 14  &  $\mathbf{326}$      & 10          \\
Fermilab--IceCube & 657 & 14  &  $\mathbf{328}$      & 10      \\
\hline
\end{tabular}
\end{footnotesize}
\end{table}

\begin{figure}
\includegraphics[angle=0,scale=.15]{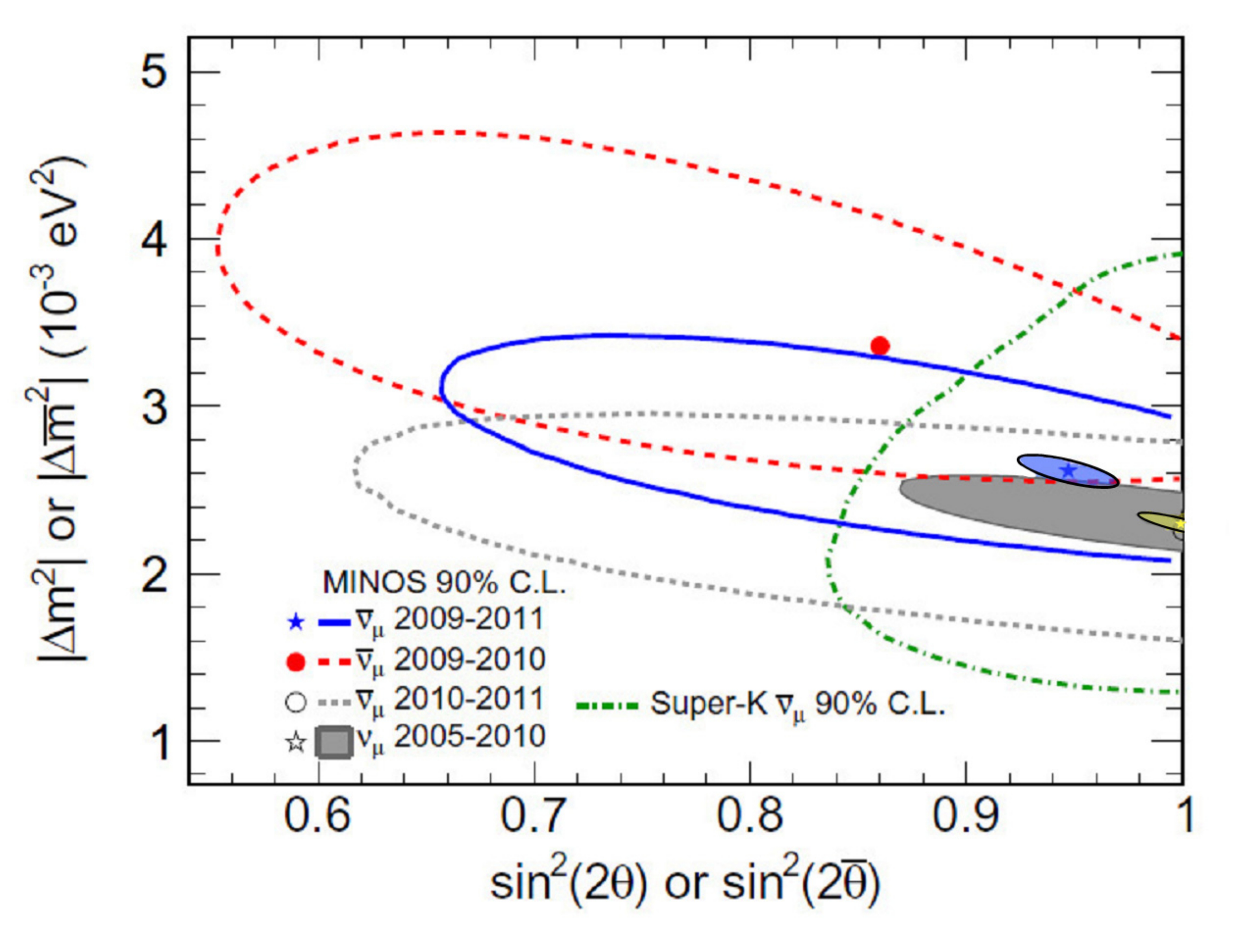}
\caption{ Neutrino and anti-neutrino muon  oscillation probability in $\sin(2 \theta_{23})=1$ and  $\Delta m^{2}=(2.35^{+0.11}_{-0.08})\cdot 10^{-3}eV$ as in old data by MINOS and SK. Also the old
 $\bar{\nu}_{\mu}$  oscillation probability  (\cite{1}) into $\bar{\nu}_{\tau}$ CPT violated parameters was
   $\Delta \overline{m}_{23}^{2} =(3.36^{+0.45}_{-0.40} )\cdot 10^{-3}$ eV$^{2}$ and $\sin^{2}(2\overline{\theta}_{23})\,=\,0.86\,\pm\,0.11$.
   On the contrary, the recent parameters are more comparable with the CPT conserved ones \cite{111}: $\Delta \overline{m}_{23}^{2} =2.62^{+0.31}_{-0.28}\cdot 10^{-3}$ eV$^{2}$, $\sin^{2}(2\overline{\theta}_{23})\,=\,0.945$.
  The early MINOS discordance was about $2.5$ sigma, but the most recent one is  within one sigma consistent with the CPT conserved case. Our beaming across the Earth might reach a discrimination described somehow by the inner smaller ellipses, whose extension at 6 sigma may disentangle even last eventual MINOS tiny CPT discordance. }\label{MINOS-ellipse}
\end{figure}

\section{Conclusion}
Our recent proposal, \emph{NeuWorld} experiment \cite{-1} shows the  estimated  ${\nu}_{\mu}$ ${\nu}_{\tau}$ event rates  (and the same for the anti-particles)
for two main neutrino baselines and configurations  (CERN (or Fermilab) to ICECUBE); the experiment is offering a high rate of detection of muon disappearance in any MINOS like CPT violation scenario,
as well as a sharp probe of
the appearance of many $\tau$ and $\bar{\tau}$: at least two or one at a day. This \emph{NeuWorld} experiment may lead  also to an improvement
of many neutrino oscillation parameter measure \cite{-1} not discussed here, as the neutrino mass hierarchy.
Such up-going neutrino beam signals in DeepCore and SuperK is completely noise free from any atmospheric neutrino ones,
since the clock and packet time-windows of the sent bunches is narrow and the directionality is also  very selective. The \emph{NeuWorld} experiment
is a very promising, being a very low cost one while  at once offering the largest neutrino baseline test, beaming from CERN (or Fermilab) to largest neutrino detector (Deep Core) that  already exist:
 a bending tunnel of a few percent OPERA size is just the missing tool.

\end{document}